\documentclass[aps,prl,showpacs,floatfix,reprint,superscriptaddress]{revtex4-1}
\usepackage{graphicx} 
\usepackage{hyperref}
\usepackage{amsmath}
\usepackage{amssymb}

\newcommand{\BaCo}{Ba(Fe$_{1-x}$Co$_x$)$_2$As$_2$}
\newcommand{\deltaRff}{$\Delta\rho_{\rm ff}$/$ \rho _{\rm n}$~}

\newcommand{\Rff}{$\rho_{\rm ff}$~}

\begin{document}

\title{Universality and unconventional enhancement of flux-flow resistivity in \BaCo}

\author{X. Y. Huang}
\affiliation{Department of Physics, Kent State University, Kent, Ohio, 44242, USA}

\author{D. J. Haney}
\affiliation{Department of Physics, Kent State University, Kent, Ohio, 44242, USA}

\author{Y. P. Singh}
\altaffiliation[Present Address: ]{Department of Mechanical Engineering, The University of Akron, Akron, Ohio, 44325, USA}
\affiliation{Department of Physics, Kent State University, Kent, Ohio, 44242, USA}

\author{T. Hu}
\affiliation{Department of Physics, Kent State University, Kent, Ohio, 44242, USA}
\affiliation{Shanghai Institute of Microsystem and Information Technology, Shanghai 200050, China}

\author{H. Xiao}
\affiliation{Department of Physics, Kent State University, Kent, Ohio, 44242, USA}
\affiliation{Center for High Pressure Science and Technology Advanced Research, Beijing 100094, China}

\author{Hai-Hu Wen}
\affiliation{Nanjing University, Nanjing 210093, China}

\author{M. Dzero}
\affiliation{Department of Physics, Kent State University, Kent, Ohio, 44242, USA}

\author{C. C. Almasan}
\affiliation{Department of Physics, Kent State University, Kent, Ohio, 44242, USA}

\date{\today}

\begin{abstract}
Measurements of the current-voltage characteristics ($I$-$V$) were performed on \BaCo~single crystals with doping level $0.044 \leq x \leq 0.1$. An unconventional increase in the flux-flow resistivity \Rff with decreasing magnetic field $H$ was observed across this doping range. Such an abnormal field dependence of \Rff is in contrast with the linear $\rho_{\rm ff}(H)$ of conventional type-II superconductors, but similar to the behavior recently observed in the heavy-fermion superconductor CeCoIn$_5$. A significantly enhanced \Rff was found for the $x=0.06$ single crystals, implying a strong single-particle energy dissipation around the vortex cores. At different temperatures and fields and for a given doping concentration, the normalized \Rff scales with normalized field and temperature. The doping level dependence of these parameters strongly suggests that the abnormal upturn in \Rff is likely related to the enhancement of spin fluctuations around the vortex cores of the samples 
with $x\approx0.06$.
\end{abstract}

\maketitle

\section{Introduction}
The cobalt-doped superconducting (SC) iron-arsenide material \BaCo~ 
has been widely studied  due to the presence of a magnetic phase transition above the zero-field superconducting critical temperature T$_{c0}$ \cite{FernandesPRB2010,RotunduPRB2011Secondorder,ChuPRB2009,FernandesPRB2010} and the microscopic co-existence of magnetic and superconducting phases under the superconducting dome \cite{FernandesPRB2010}. The latter finding also provides the motivation to investigate whether these phases may be governed by the system's proximity to a putative quantum critical point (QCP) \cite{FanlongNingJPSJ2009,fernandesPRL2013howmanyQCP,ThermPowerQCP}. Recently, the co-existence of superconductivity and spin-density wave (SDW) phases has been further confirmed by NMR measurements, in which a spin glass phase is revealed between $x=0.06$ and $0.071$ \cite{CurroPRL2013}. Intriguingly, within the framework of multiband theory of superconductivity, the doping-induced disorder seems to play a major role in suppressing magnetic order while giving rise to emergent superconducting order since the superconductivity remains immune to the intraband disorder-induced scattering processes while SDW order does not  \cite{VC_Disorder2011}. 
\begin{figure}
\centering
\includegraphics[width=1\linewidth]{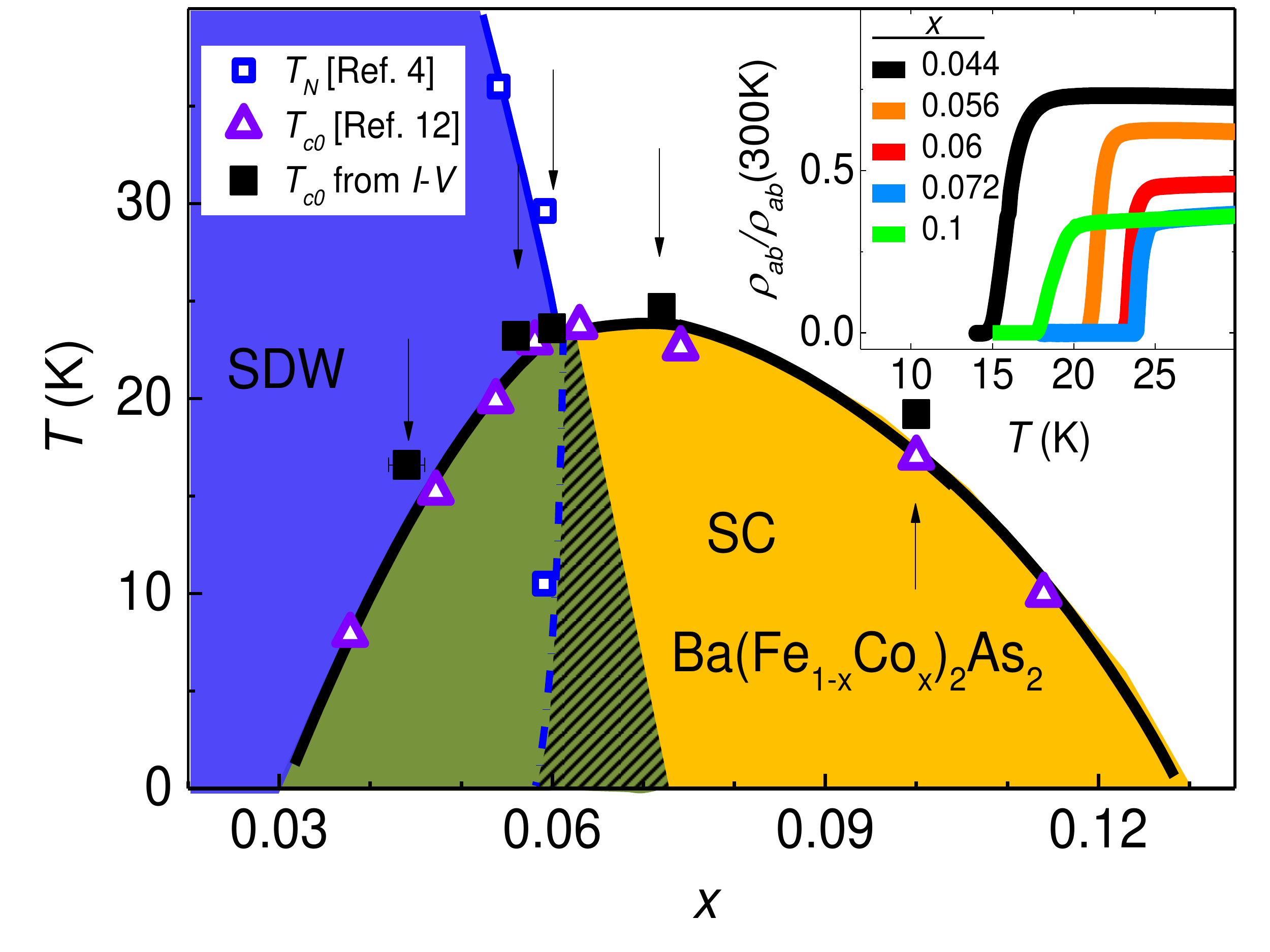}
\caption{\label{f1}Sketched of the temperature - doping $T$-$x$ phase diagram for \BaCo. The phases depicted on this phase diagram are: SDW (blue region), coexistence of SC and SDW phases  \cite{FernandesPRB2010} (green region), spin glass phase \cite{CurroPRL2013} (hatched green region), and pure SC phase (yellow region).  The arrows mark the five doping levels discussed in this paper. Inset:  In-plane resistivity normalized to its value at 300 K, $\rho/\rho(300$ K), measured in zero field and for five doping levels.}
\end{figure}
 
Flux-flow resistivity \Rff measurements offer a crucial insight into competing interactions of a superconducting system since 
the energy dissipation due to the motion of the flux vortices in the Ohmic regime is sensitive to the fluctuations of the corresponding order parameters \cite{TaoPRL}, either of magnetic or some other origin. In these measurements one traces the evolution of \Rff (the slopes in the current-voltage $I$-$V$ characteristics) with magnetic field and temperature and, in principle, extracts information about the band structure \cite{Silaev2016} as well as relevance of various dissipation mechanisms to the transport properties of flux vortices. The idea behind the experiment \cite{TaoPRL} originates from the following observation: the dissipation in the form of flux-flow resistivity is strongly affected by the magnetic fluctuations around the vortex cores of an unconventional superconducting system in the proximity to a magnetic instability. 

In this paper, we use our improved $I$-$V$ measurement method to probe the physics related to the co-existence of magnetic and superconducting phases in \BaCo~(see Fig. \ref{f1}) by exploring the vortex behavior that could be affected by the secondary (in relation to superconductivity) phases in the mixed state. We find that an abnormal flux-flow resistivity as a function of magnetic field appears over a wide range of doping. Intriguingly, we find that the $x = 0.06$ single crystal shows the largest upturn caused by the strongest dissipation among the five doping levels we have studied. We naturally interpret this observation as an indication that the secondary phase has the strongest effect for the $x\approx 0.06$ single crystals. In addition, we reveal a universal scaling behavior of the flux-flow resistivity data obtained at different magnetic fields and temperatures for different Co doping for this \BaCo~ system (see below). We investigate how the parameters which enter this scaling function change with doping.  While the scaling form of the flux-flow resistivity with magnetic field is universal for the whole doping range studied, the variation of the other two scaling parameters allows us to identify three regimes with distinctive dependence of the flux-flow resistivity on doping and temperature. All these results point to a strong magnetic field $H$ and temperature $T$ dependence of the viscosity coefficient $\eta$ as a result of the system's proximity to a magnetic instability.
 
\section{Experimental Details}
Single crystals of \BaCo~were grown using FeAs self-flux method \cite{ChuPRB2009,NiNiPRB2008EffectofCo}. The actual Co-doping level $x$ of each single crystal was determined by comparing its zero-field superconducting transition temperature $T_{c0}$ value to well-established $T_{c0}$-$x$ phase diagrams for this system \cite{NiNiPRB2008EffectofCo,ChuPRB2009,J.Reid_PRB2010}. As shown in Fig. \ref{f1}, the doping levels of the single crystals discussed in this paper are $x=$0.044, 0.056, 0.06, 0.072, and 0.1, which cover the under-doped, optimally-doped, and over-doped regimes. The temperature and magnetic field dependence of the electrical resistivity $\rho$ was measured on thin single crystals using the standard four-probe method with current flowing in the $ab$ plane and $H$ applied along the crystallographic $c$ axis. The inset to Fig.~\ref{f1} shows that all the single crystals studied here have sharp superconducting transitions. 

We also performed $I$-$V$ measurements as a function of $T$ and $H$. Due to strong vortex pinning present in the mixed state of these superconductors, we had to apply a large current ($I \leq 160$ mA).  In order to maximize the current density (desirable for the effective de-pinning of the flux vortices), the cross-section area $A$ of the single crystals was reduced down to $0.17\times0.04$ mm$^2$; indeed, for a given heating power per unit length $l$, a maximum current density $j$ is accomplished for an achievable minimum cross-section area $A$ since $l\equiv(I^2R)/L=(j^2A^2)\cdot(\rho L/A)/L= j^2\rho A$. 

\begin{figure}
\centering
\includegraphics[width=1\linewidth]{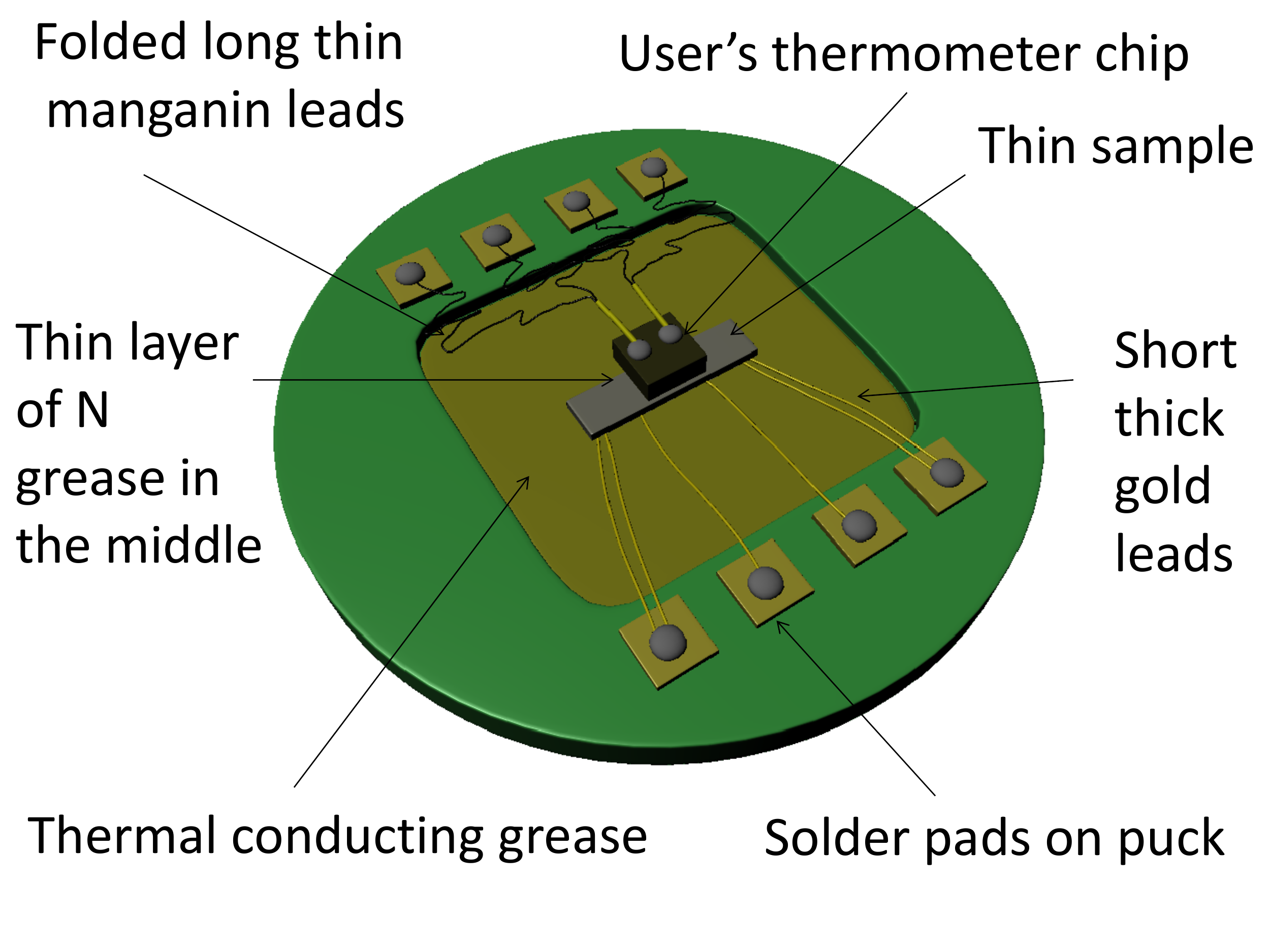}
\caption{\label{f2}
Schematic drawing of the current - voltage $I$-$V$ measurement configuration that takes care of Joule heating caused by the large applied current. }
\end{figure}

To minimize the Joule heating of the gold current leads and also to increase the heat transport from the single crystal to the thermal bath, multiple short thick gold current leads were used for the two current terminals (see Fig. \ref{f2})  since for a given applied current, the dissipated power $ P= I^2R = I^2\rho L/A$. Also, we used Sn, instead of silver paste, in order to decrease the contact resistance between the single crystal and the current leads down to less than $10$ $\mu\Omega$ \cite{TanatarIOP2010}. 

In order to increase the temperature stability and to be able to apply the large current values, we have also improved our measuring protocol and apparatus. We added an additional thermometer, mounted on the top of the sample using $N$-type grease, which we used to control and measure the temperature of the sample, and we used long folded manganin wires as terminal leads of this additional thermometer in order to decrease the heat transport between the thermometer and puck, since manganin has poor thermal conduction.  After the temperature was deemed stable, a 60 sec wait time (with the persistent current flowing through the single crystal) was included into the measurement sequence and only then the $I$-$V$ data were collected. Finally, we used  a combination of Linear Research  ac resistance bridge LR700 with extended current limit and Physical Property Measurement System (PPMS) to carry out the $I$-$V$ measurements. Due to the high current limit imposed by our experimental conditions ($I \approx 160$ mA) and the fact that vortex pinning increases with decreasing temperature, all the $I$-$V$ measurements were done at temperatures 0.87 $T_{c0}\leq T\leq$ 0.98 $T_{c0}$, i.e. about 2 K below the $H$-$T$ phase boundary. Using all the improvements in our experimental technique mentioned above, we were able to reduce the temperature instability due to Joule heating to less than 0.1 K.
  
\section{Results}
\subsection{Current-voltage characteristics} The electrical resistivity in the mixed state of type-II superconductors in the presence of an applied magnetic field is mainly governed by the motion of Abrikosov vortices \cite{Y.B.Kim1965}. When the Lorentz force is larger than the pinning force, the flux vortices are driven into a viscous-flow state. The flux-flow resistivity \Rff is defined as \Rff$\equiv k\cdot dV/dI$, where $dV/dI$ is the slope of the linear region of the $I$-$V$ curves and $k$ is a geometric factor. The intercept obtained by extrapolating the linear $V(I)$ region to zero voltage gives the value of the critical current $I_c$. The flux-flow is solely determined by the bulk properties of the material. Therefore, the $I$-$V$ measurement under the superconducting dome is a probe that takes advantage of the fact that an external magnetic field induces vortices in type-II superconductors that form islands of dissipative matter embedded in the non-dissipative superconductor and allows one to measure the dissipation of these flux vortices. Consequently, the application of this technique makes it possible to probe the interaction between superconducting and normal regions of a sample at the length scale determined by the vortex size, as flux vortices move through the superconductor. 

\begin{figure}
\centering
\includegraphics[width=1\linewidth]{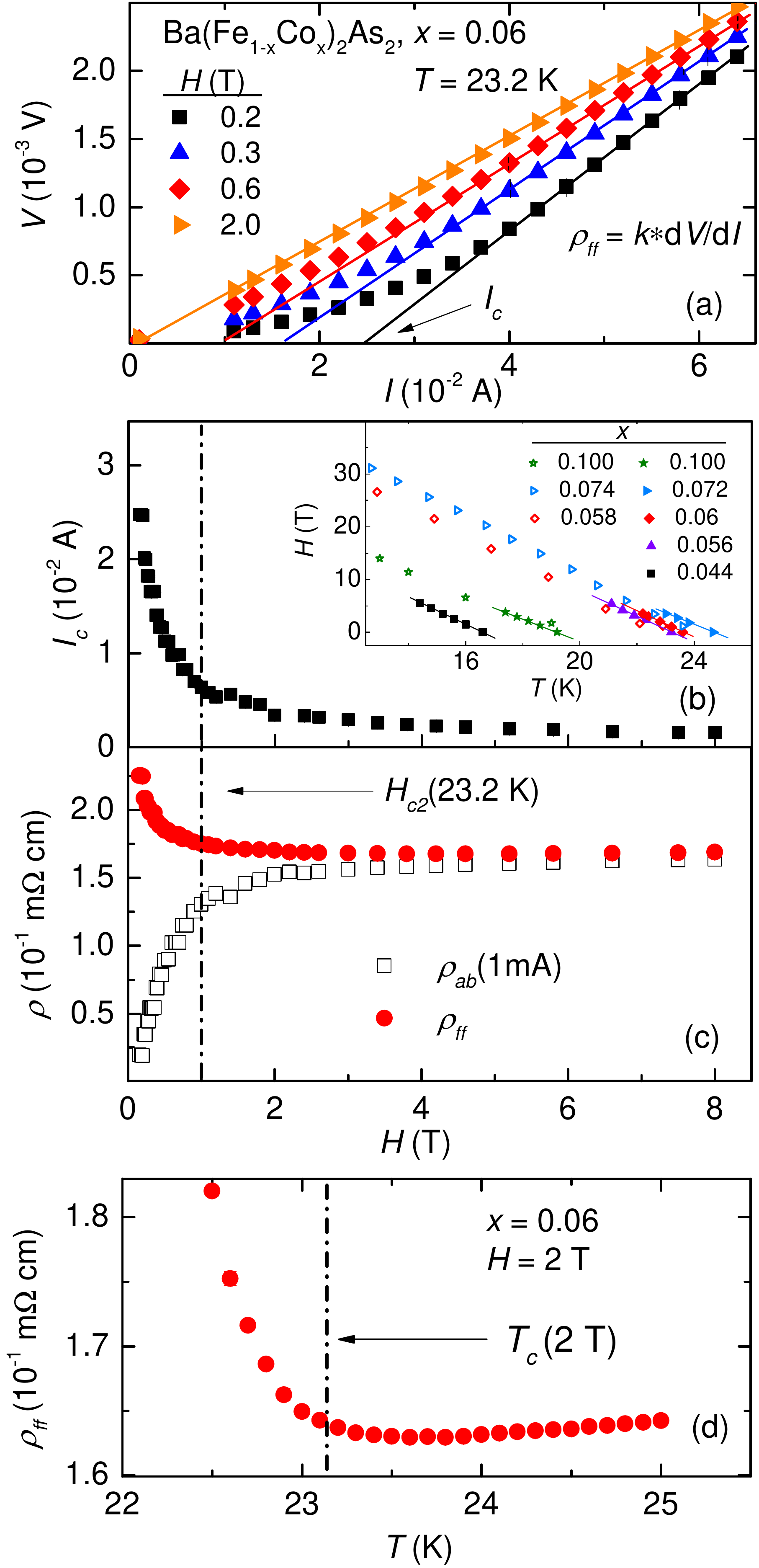}
\caption{\label{f3}
(a) Typical current-voltage $I$-$V$ curves measured in the mixed state of \BaCo~at a temperature $T$ of 23.2 K and different magnetic field $H$ values. (b) Plot of the critical current $I_c$ vs $H$. Inset to (b):  $H$-$T$ phase diagram obtained from this work (filled symbols) and from \cite{NiNiPRB2008EffectofCo} (open symbols). (c) $H$ dependence of the flux-flow resistivity \Rff (filled circles) and the resistivity $\rho_{ab}$ measured with a constant current of 1 mA applied parallel with the $ab$ plane of the single crystal (open squares). (d) \Rff vs $T$ measured at 2 T on the $x=0.06$ single crystal.}
\end{figure}

Figure \ref{f3}(a) shows $I$-$V$ data measured in the mixed state of \BaCo~with $x=$ 0.06 at $T=23.2$ K $=0.98~T_{c0}$ and different field values. These $I$-$V$ characteristics are typical for all the samples studied. The lines show the slopes of the Ohmic regime measured for different $H$
values, which allow us to determine $\rho_{\rm ff}(H)$. Notice that the linear region (Ohmic regime) becomes wider with increasing $H$ and that it extends over the whole current range (the $I$-$V$ characteristics are straight lines passing through origin) at $H$ values equal to or larger than the upper critical field $H_{c2}(T)$, reflecting the Ohmic behavior of the sample in the normal state.

Figure~ \ref{f3}(b) is a plot of $I_c(H)$  at $T=23.2$ K, extracted from the $I$-$V$ data measured at this temperature as discussed above. Notice that $ I_c$ shows a sharp increase with decreasing field below $H \approx1$ T. We define this field value below which the critical current increases abruptly as the upper critical field $H_{c2}$ corresponding to this temperature.
The inset to Fig. \ref{f3}(b) shows for the Co doping studied here the $H$-$T$ phase diagram generated as just discussed (filled symbols) and from  published resistivity data (open symbols) in which $H_{c2}(T)$ is defined as the onset in the SC transition in $\rho(H)$ curves measured at multiple temperatures \cite{NiNiPRB2008EffectofCo}. Notice the good agreement between our results and published results, supporting our definition of $H_{c2}(T)$ shown in Fig.~\ref{f3}(b).

Figure \ref{f3}(c) shows the field dependence of \Rff (red filled circles) extracted from the $I$-$V$ curves shown in Fig. \ref{f3}(a), along with the resistivity $\rho_{ab}$ measured using a low constant current of 1 mA applied in the $ab$ plane of the crystal (open squares). The difference between these two curves at low $H$ values is a result of the fact that the data shown in red circles give the flux-flow dissipation of the vortices in the Ohmic regime where pinning is negligible, while the data shown as squares give the dissipation of the vortices in the non-Ohmic regime where pinning dominates. As expected, in the normal-state (when $H>H_{c2}$) both sets of data corresponding to the two measurements overlap within around 4\%, and both \Rff and $\rho_{ab}$ show a weak field dependence, reflecting the weak magnetoresistance of this sample. Also, notice that \Rff increases sharply, while $\rho_{ab}$ displays a sharp decrease for $H<H_{c2}=1$ T corresponding to this temperature. 

Type-II superconductors display a well known linear relationship between \Rff and $H$ at low $H$ values, with  $\rho_{\rm ff}/\rho_{\rm n}\propto H/H_{c2}$, where $\rho_n \equiv \rho_{\rm ff}(H_{c2})$ \cite{BardeenStephen1965}, and its saturation near $H_{c2}$  \cite{KopninPRB1995,KoppinPRL1997,KambePRL1999UPt3}. Indeed, for $H \approx H_{c2}$ the \Rff  data approach the $\rho_n$ data, as shown in Fig.~\ref{f3}(c), which is consistent with the fact that \Rff~$\approx \rho_n$ for $H \approx H_{c2}$ \cite{Y.B.Kim1965}. However, the upturn of \Rff vs $H$ revealed by the data of Fig.~\ref{f3}(c) in the mixed state is in contrast with this well known behavior. Since the dissipation of the vortices is dominated by the dissipation of the quasiparticle in the vortex cores at least for $T\sim T_c$, the upturn in  \Rff in the mixed state reflects the increase in the scattering of the quasiparticles in the vortex cores with decreasing applied magnetic field. Hence, the dissipation of the vortices reveals the dominant scattering mechanism of the underlying normal state. 

Figure \ref{f3}(d) shows the temperature dependence of \Rff extracted from the $I$-$V$ curves measured at a constant field of 2 T for the $x=0.06$ single crystal. Notice the sharp increase of \Rff with decreasing $T$ just below $T_c \approx 23.1$ K for this value of the applied magnetic field. This behavior is typical for all the samples studied. We note that this nonmetallic vortex dissipation displayed in this figure is in sharp contrast with the metallic dissipation in the normal state. One would expect the scattering of the quasiparticles of the vortex cores and normal state to be very similar to each other near $T_c$.  However, possible deviations from this behavior could occur due to the presence of several competing interactions, as discussed below. 

\begin{figure}
\centering
\includegraphics[width=1.0\linewidth]{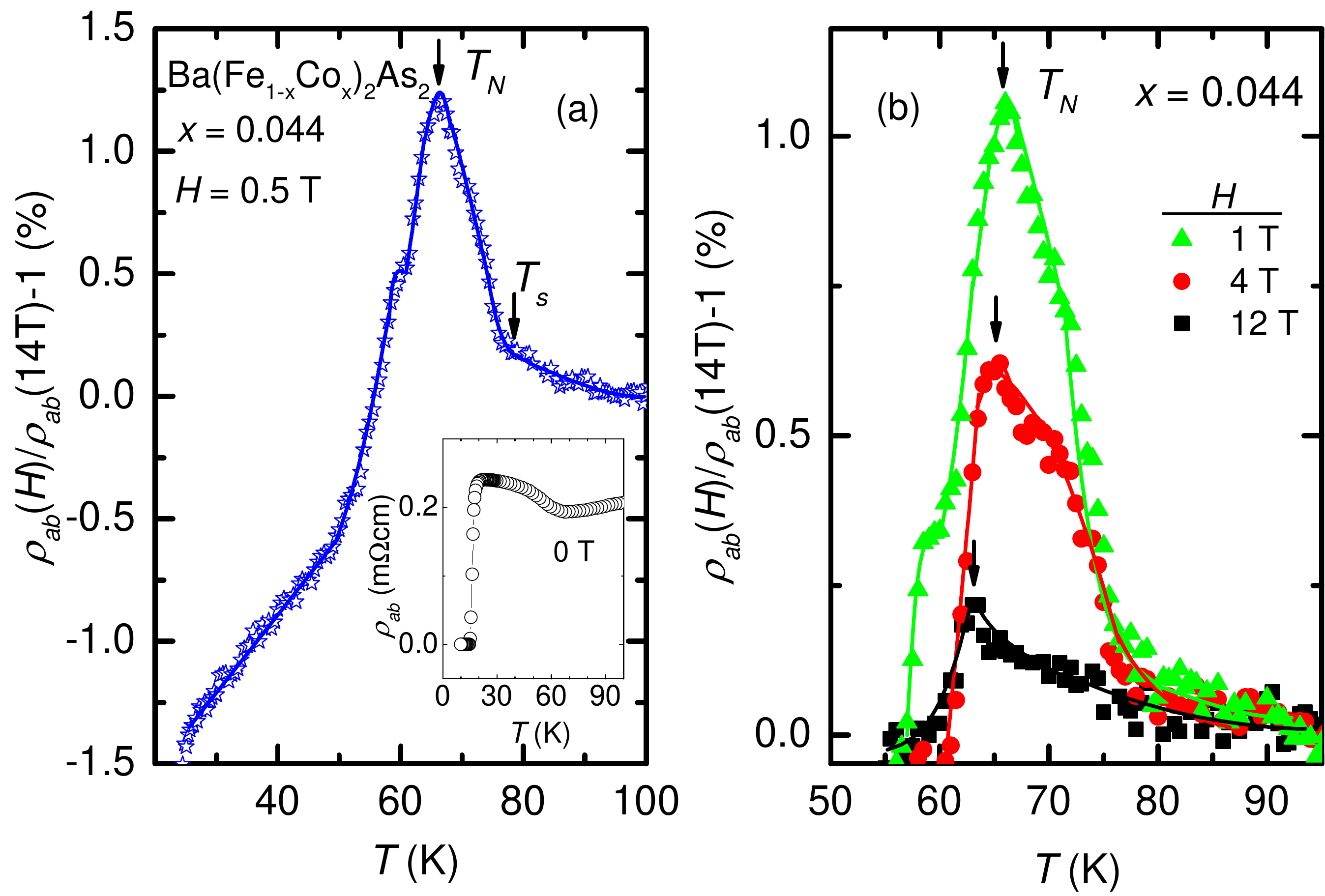}
\caption{\label{f4} (a) Temperature $T$ dependent magnetoresistivity $MR \equiv \rho_{ab}(H)/\rho_{ab}(14$ T$)-1$ curve of the $x=0.044$ single crystal measured in a magnetic field of 0.5 T and with a current of 1 mA. Inset: $\rho_{ab}$ vs $T$ measured with no magnetic field applied. (b) $T$ dependent MR curve for 1, 4, and 12 T.}
\end{figure}

It is well known that the scattering of quasiparticles is enhanced by critical spin fluctuations present close to a magnetic transition.  For example, enhanced electrical resistivity due to magnetic spin fluctuations has being reported just above the antiferromagnetic phase transition in the normal state of CeCo(In$_{1-x}$Cd$_x$)$_5$ with $x=0.0075$ \cite{NairPNAS2010}. In addition, our magnetoresistivity $MR \equiv \rho(H)/\rho(14$ T)$-1$ data of the under-doped Ba(Fe$_{1-x}$Co$_x$)$_2$As$_2$ ($x = 0.044$), with a Neel temperature $T_N = 66.3$ K  \cite{D.K.PrattPRL2009} larger than $T_{c0}=16.6$ K, show as $T$ decreases 
first a sudden change in slope followed by a peak [Fig.~\ref{f4}(a)]. We identify the structural and magnetic phase transitions, as
$T_s$ = 76.6 K and $T_N$ = 66.3 K, respectively, corresponding to these two features in the MR curve, as indicated by the arrows on the figure. These transition temperatures agree with published data of heat-capacity, susceptibility, resistivity, Hall-coefficient, and neutron diffraction measurements on the same doping \cite {ChuPRB2009,NiNiPRB2008EffectofCo,FernandesPRB2010}. The maximum MR at $T_N$ reflects the maximum quasiparticle scattering due to critical magnetic fluctuations present near $T_N$ \cite{TaoPRL}.  Hence, quasiparticle scattering identifies the presence of critical magnetic fluctuations for systems that are close to a magnetic  instability, in this case SDW.  

Typical MR data in applied magnetic fields of 1, 4, and 12 T are shown in Fig.~\ref{f4}(b).  The position of the maximum in MR shifts to lower temperatures with increasing $H$, confirming that, indeed, the position of the maximum in MR represents $T_N$. 
We note that in plotting MR, we effectively subtracted the background scattering present in a magnetic field of 14 T (at least for $T\geq 60 $ K) since $T_N<60$ K for this $H$ value.  As a result, we are able to extract the quasiparticle scattering due to spin fluctuations in the vicinity of the SDW order from the total quasiparticle scattering - information unrevealed by the direct measurement of $\rho(T)$ [inset to Fig. \ref{f4}(a)]. 

Based on the above discussion, we are led to interpret the observed upturn in $\rho_{\rm ff}(T)$ with decreasing $T$ [Fig. ~\ref{f3}(c)] as being due to critical antiferromagnetic fluctuations in the vicinity of the boundary separating antiferromagnetic and paramagnetic phases. Also, since these magnetic fluctuations are suppressed by a magnetic field, $\rho_{\rm ff}$ is strongly suppressed, as expected, with increasing $H$ [see Fig. \ref{f3}(c)]. The fact that  \Rff starts increasing just below the SC boundary [Figs. \ref{f3}(c) and \ref{f3}(d)] suggests that the dynamic AFM fluctuations emerge at the SC phase boundary.

We note that the well-known positive slope in $\rho_{\rm ff}(H)$ was observed in some iron-based superconductors like  LiFeAs \cite{MaedaPRB2012}, NaFe$_{0.97}$Co$_{0.03}$As \cite{MaedaPhyC2013}, BaFe$_2$(As$_{0.55}$P$_{0.45}$)$_2$ \cite{MaedaPhyC2014}, and FeSe$_{0.4}$Te$_{0.6}$ \cite{MaedaPRB2015} using a microwave technique. Nevertheless, all these flux-flow studies are done at low temperatures, while our $I$-$V$ measurements are limited to high temperatures, a few degrees below $T_{c0}$.  Therefore, the difference between the published $\rho_{\rm ff}(H)$ dependence (measured at low-temperatures) and the present data  (measured at temperatures close to $T_{c0}$) reflects the difference in the scattering mechanism of the quasiparticles in the vortex cores  in these two different temperature regimes. One such difference could be that the low $T$ published data are not affected by critical spin fluctuations.

\subsection{Doping dependence of flux-flow resistivity} 
\begin{figure}
\centering
\includegraphics[width=1.0\linewidth]{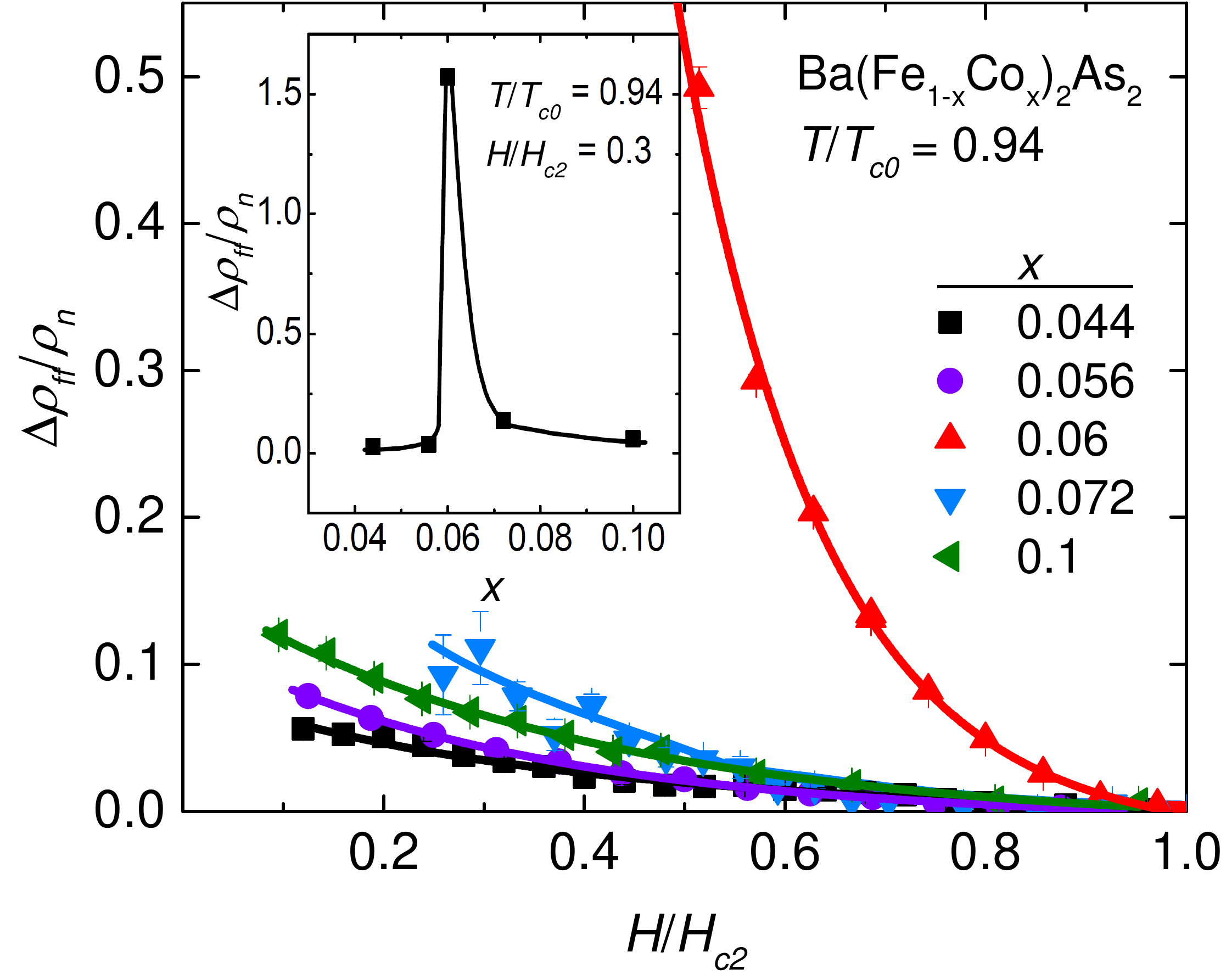}
\caption{\label{f5} Normalized flux-flow resistivity $\Delta \rho_{\rm ff}/\rho_{\rm n} \equiv \rho_{\rm ff}/\rho_{\rm n}-1$ as a function of reduced magnetic field $H/H_{c2}$ measured at the same reduced temperature $T/T_c=0.94$ for all Co doping values $x$. Inset: $\Delta\rho_{\rm ff}/\rho_n$ vs $x$ measured at $T/T_c=0.94$ and $H/H_{c2}=0.3$ for all the single crystals studied.}
\end{figure}

On Fig. \ref{f5} we plot the normalized flux-flow resistivity $\Delta \rho_{\rm ff}/\rho_{\rm n} \equiv \rho_{\rm ff}/\rho_{\rm n}-1$ as a function of reduced field $H/H_{c2}$  measured at the same reduced temperature $T/T_{c0}=0.94$ for all doping levels. Notice that $\Delta \rho_{\rm ff}/\rho_{\rm n}$ displays a weak field dependence and a weak change in its value as the doping level  increases from $x=0.044$ (black squares) to $x=0.056$ (purple circles). However, $\Delta \rho_{\rm ff}/\rho_{\rm n}$ reveals a huge upturn for the $x=0.06$ single crystal (red triangles), a doping level around optimal-doping (see Fig. \ref{f1}). With further increasing the doping level to the over-doped regime, the upturn decreases to values as low as the ones found in the $x=0.044$ single crystals. 

The fact that the strongest upturn in $\Delta\rho_{\rm ff}/\rho_{\rm n}$ happens at or around optimal doping while it remains rather weak in the underdoped and overdoped regimes indicates a significant increase in the scattering of the quasiparticles around the vortex cores for the $x \approx 0.06$ single crystals. This, in turn, suggests that the critical spin fluctuations are the strongest for the optimally-doped Co single crystal, providing a leading contribution to the energy dissipation of the moving flux vortices; hence, it further suggests that the $x=0.06$ Co doping is right at the phase boundary between a magnetically ordered state (SDW) and a spin disordered state (paramagnetic phase). Indeed, this particular Co doping is the closest, amongst the doping studied here, to the doping where the SDW phase boundary enters under the SC dome (see Fig.~\ref{f1}). In the underdoped regime of the  \BaCo~system, the SDW order state coexists with the SC state. Thus, the spin fluctuations around vortex cores should be considerably suppressed, which is consistent with the weak upturn observed in $x=0.044$ and 0.056. The weak upturn in $\Delta \rho_{\rm ff}$ with decreasing $H$ in the overdoped regime could be a result of the fact that these samples are further away from the SDW phase boundary. 

We note that we can exclude the disorder effect induced by the Co doping as the dominant scattering source that gives rise to the enhanced quasiparticle dissipation, since the dependence of the upturn in $\Delta \rho_{\rm ff}/\rho_{\rm n}$  on the doping-level is non-monotonic and since the $x=0.06$ samples is most likely the least disordered, having the highest $T_{c0}$. 
 
\begin{figure}
\centering
\includegraphics[width=1.0\linewidth]{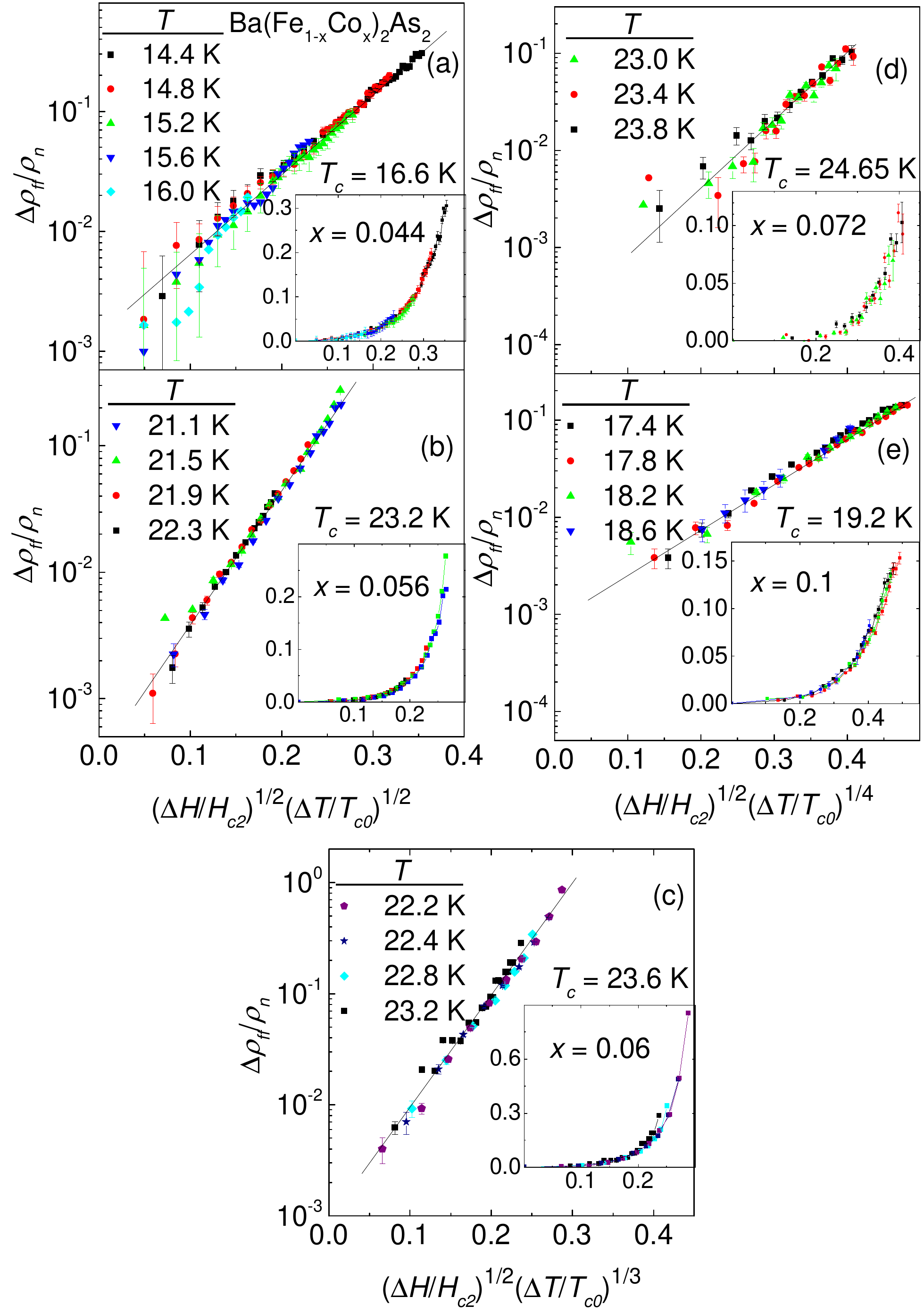}
\caption{\label{f6} Semi-log plots of the normalized free-flux-flow resistivity $\Delta \rho_{\rm ff}/\rho_{\rm n}$ vs reduced field and temperature $(\Delta H/H_{c2})^{m}(\Delta T/T_{c0})^{n}$ for all five doping levels of \BaCo studied. The straight lines are guides to the eye. Inset: Linear plots of the same data. 
}\end{figure} 

\subsection{Scaling behavior}

\begin{figure}
\centering
\includegraphics[width=1.0\linewidth]{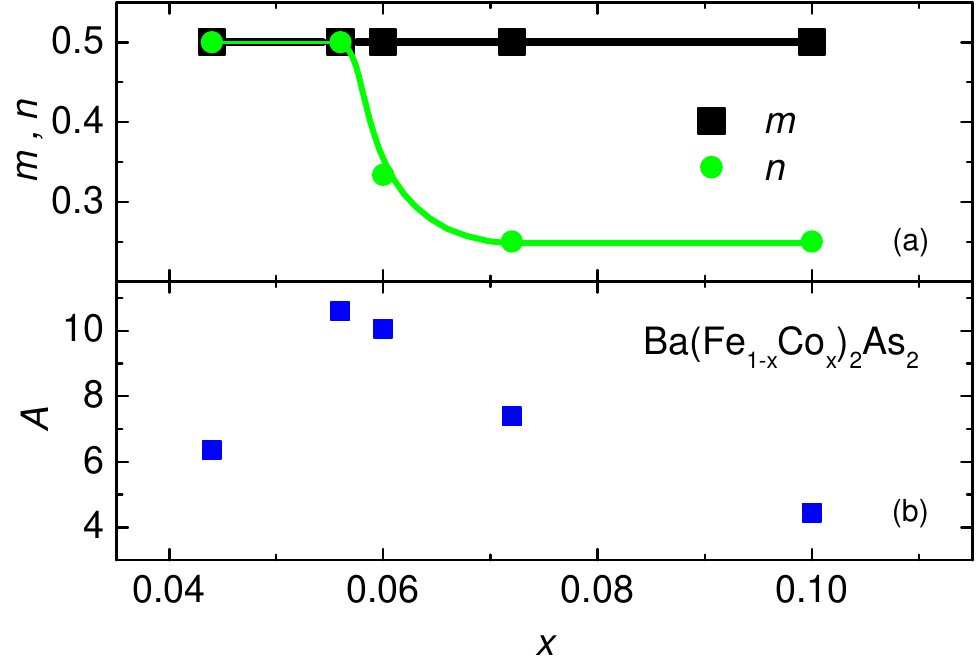} 
\caption{
\label{f7}Doping dependence of the scaling parameters $m$ and $n$ (top panel), and $A$ (bottom panel). 
}

\end{figure}
A careful analysis of all the $I$-$V$ data obtained in different applied magnetic fields, temperatures, and for different Co doping for the \BaCo~ system reveals a universal scaling behavior between normalized flux-flow resistivity $\Delta \rho_{\rm ff}/ \rho _{\rm n}$ and reduced field $\Delta H/H_{c2}$ and temperature $\Delta T/T_{c0}$. Figures \ref{f6}(a) through \ref{f6}(e) are semi-log plots of \deltaRff as a function of $(\Delta H/H_{c2})^{m}(\Delta T/T_{c0})^{n}$ for the five doping levels studied here, while their insets are plots of the same data on a linear scale. Indeed, all these plots show that all the $\Delta\rho_{\rm ff}$ data for the same doping obtained at different fields and temperatures scale for certain values of the scaling parameters $m$ and $n$, while the linear correlation on the semi-log plot between $\Delta\rho_{\rm ff}$ and $(\Delta H/H_{c2})^{m}(\Delta T/T_{c0})^{n}$ reveals an exponential functional dependence. Hence, all the data follow the functional form:
\begin{eqnarray}
\label{e1}
{\rm log}_{10}( C\Delta\rho_{\rm ff}/ \rho _{\rm n}) = A(\Delta H/H_{c2})^m(\Delta T/T_{c0})^n,
\end{eqnarray}
where $C$ is a constant, $\Delta\rho_{\rm ff} \equiv \rho_{\rm ff}-\rho_{\rm n}$, $\Delta H \equiv H_{c2}-H$, and $\Delta T \equiv T_{c0}-T$, while $A(x)$, $m(x)$, and $n(x)$ are mutually independent parameters that only depend on the doping level $x$. The $H_{c2}$ and $T_{c0}$ values used for the different doping levels were obtained as explained in the discussion related to Fig. \ref{f3}(b) and are shown in the $H$-$T$ phase diagram of the inset to this figure. 

The scaling parameters $m$, $n$, and $A$ are plotted as a function of the doping $x$ in Figs. \ref{f7}(a) and \ref{f7}(b). These figures show that the field dependence of the flux-flow resistivity is the same, $m=0.5$, across the whole doping range, while the temperature exponent $n$ displays a step-like dependence on $x$ with $n=1/2$, 1/3, and 1/4 for the under-doped, optimally-doped, and over-doped range, respectively [see Fig. \ref{f7}(a)]. The  scaling parameter $A$ displays a peak for $x\approx0.06$. The doping dependence of $n$ and $A$ is in agreement with the sharp increase in vortex dissipation for the $x=$0.06 single crystal  [see Fig. \ref{f5}].

The fact that no doping dependence is found in $m$ implies that the unconventional behavior in $\rho_{\rm ff}(H)$ is caused by the the same scattering mechanism of the quasiparticles around the vortex cores for samples belonging to the three doping regimes. On the other hand, $n$ and $A$ are most likely related to the magnetic degrees of freedom and their effect on the scattering of the flux vortices. Although further theoretical study is needed to understand more about this scaling behavior of the quasiparticle dissipation in the  \BaCo~system, this $I$-$V$ study strongly supports the change of the spin dynamics around the vortex core for the $x\approx0.06$ samples. 

\subsection{Discussion}
As we have mentioned above, the motion of the flux vortices is determined by the balance of three forces: (i) Lorenz force 
${\vec f}_L\propto {\vec j}\times{\vec B}$, where ${\vec j}$ is the transport current passing through the superconductor and $\vec B$ is the magnetic induction, (ii) pinning force ${\vec f}_{\textrm{pin}}$, which is due to the interaction between pinning centers in the sample and flux vortices, and (iii) viscosity force ${\vec f}_v=-\eta {\vec v}_L$, where $\eta(H,T,x)$ is the viscosity coefficient characterizing the bulk superconducting properties of the material and $v_L$ is the velocity of the flux vortices \cite{Strnad1964,Larkin1986}.  The pinning force determines the value of the critical current density $j_c$ and when the transport current $j>j_c$ the pinning force decreases with increasing vortex velocity. Thus, the only source of dissipation in the flux-flow regime is determined by the motion of the flux vortices since the interaction effects between the flux vortices and the pinning centers can be ignored. Lastly, there are also retardation effects related to the relaxation of the superconducting order parameter, which also lead to energy dissipation at low temperatures \cite{Tinkham1964,LevKopnin1971}. It follows that the flux-flow conductivity is given by \cite{Strnad1964}:
\begin{equation}
\label{rhoff1}
\sigma_{\rm ff}=\frac{\eta(H,T,x) c^2}{\phi_0B},
\end{equation}
where $\phi_0=hc/2e$ is a quantum of flux. 

In general, the viscosity coefficient $\eta\simeq\phi_0H_{c2}/\rho_nc^2$  is essentially independent of magnetic field and temperature in {\it conventional} superconductors, which renders the flux-flow resistivity to grow linearly with magnetic field. 
In fact, the linear $B$ dependence of the flux-flow resistivity holds for superconducting alloys as well as multiband superconductors. Specifically, the problem of flux-vortex motion in multiband superconductors has been recently addressed by Silaev and Vargunin \cite{Silaev2016}, who have shown that although Eq. (\ref{rhoff1}) generally holds, the resistive properties remain non-universal and depend on the system's specifics such as density of states for each band etc. In particular, the viscosity coefficient 
is shown to be given by $\eta=\pi\hbar\sum_k\nu_{F}^{(k)}(\alpha_k+\gamma_k)$, where $\nu_{F}^{(k)}$ is the density of states of the $k$-th band and coefficients $\alpha_k$ and $\gamma_k$ are determined by the superconducting order parameter $\Delta_k(r)$ and the single particle distribution function on the $k$-th band. Thus, $\eta$ remains independent of the magnetic field at least in the limit of low temperatures,
$T\ll T_c$, and small fields, $H\ll H_{c2}$, where analytical calculations can be carried out.

In contrast with the picture discussed above, the fitting of our flux-flow resistivity data at temperatures $T\sim T_c$, summarized by Eq. (\ref{e1}), shows that the viscosity coefficient in the multiband \BaCo~ superconductors appears to be strongly field dependent. Indeed, its $T$, $H$, and $x$ dependence in the limit $h=H/H_{c2}\ll 1$ can be derived by employing the Taylor expansion as:
\begin{equation}\label{etaemp}
\frac{\eta(H,T,x)}{\eta(H_{c2},T,x)}\approx h\cdot\left(1-a_1(T,x)h+a_2(T,x)h^2\right),
\end{equation}
where ${a}_{1,2}$ are expansion coefficients that depend on temperature and Co concentration. As we have argued above, the 
increase in magnetic field should lead to the suppression of magnetic fluctuations in the bulk and, therefore, to an increase in conductivity. This in turn, based on Eqs. (\ref{rhoff1}) and (\ref{etaemp}),
implies that $a_1(T,x)\gg a_2(T,x)$ and also $a_1(T,x)>1$, so that within the experimentally relevant range of fields, the ratio of the viscosity coefficient to magnetic field, hence $\sigma_{\rm ff}$, increases with increasing $h$.  

One aspect of the flux-flow in multiband superconductors, which have not been addressed so far, 
is concerning the presence of magnetic order in the vortex cores. In this regard, it has been shown recently that SDW order can emerge inside the vortex cores in the iron-based superconductors \cite{Koshelev2015}. As a consequence, one may speculate that the onset of the SDW order inside the vortex core affects the drag force and results in the magnetic field dependence of the drag coefficient.  Indeed, both the superconducting order parameter and single-particle distribution function become dependent on the value of the SDW order parameter ${\vec M}({\vec r})$ inside the vortex cores and, therefore, one may expect that the 
fluctuations of ${\vec M}$ will affect the relaxation of the superconducting order parameter, thus contributing to energy dissipation. 

The presence of the weak anomalous increase of \Rff with decreasing $H$ and $T$ in the underdoped and overdoped regimes, despite the fact that the SDW fluctuations are fully suppressed in the bulk in both of these regions, could indicate the emergence of SDW order inside the vortex cores. Interestingly, a similar effect has been studied in CeCoIn$_5$  \cite{TaoPRL}, a completely different unconventional superconductor that belong to the '115' family of heavy fermion superconductors. In studying the magnetic fluctuations under the superconducting dome using $I$-$V$ measurements to extract flux-flow dissipation, Hu and collaborators  \cite{TaoPRL} have shown that the interplay between superconductivity and magnetism in the vortex cores is also present in CeCoIn$_5$ in which partially unscreened local moments on Ce sites show tendency towards antiferromagnetic AFM order. They identified a scaling relationship from the $\rho_{\rm ff}(T,H,P)$ data and obtain an explicit equation for the AFM boundary inside the SC dome and an AFM QCP line that is accessed with two control parameters: $H$ and $P$. However, they found a much weaker - power-law dependence - of $\rho_{\rm ff}(T,H)$ in CeCoIn$_5$, in contrast with the exponential dependence observed here in \BaCo [see Eq. (1)]. This difference is most likely due to the nature of local moments in CeCoIn$_5$ compared to the itinerant magnetism in \BaCo. In addition, Park and collaborators have also studied the interaction between superconductivity and magnetism in the heavy fermion superconductor CeRhIn$_5$ and they have revealed the presence of a field-induced QPT line under the SC dome of  that separates coexisting SC and AFM phases from a pure unconventional SC phase \cite{ParkNature2006}. 

The interplay between superconductivity and antiferromagnetism has also been studied in La$_{1.9}$Sr$_{0.1}$CuO$_4$ through neutron scattering, showing that its vortex state can be regarded as a mixture of a superconducting spin fluid and a core containing a nearly ordered SDW state \cite{LakeScience2001,LakeNature2002}.  In order to account for this result, Demler and co-workers proposed a model that assumes that the superconducting state is near a quantum phase transition (QPT) to a state with microscopic coexistence of SC and magnetic orders \cite{YZhangPRB2002}.  They have shown that when $H$ penetrates an unconventional superconductor in which the SC energy gap has nodes on the Fermi surface, field-induced quantized vortices have a magnetic ground state that suppresses superconductivity around the vortices.  The suppression of the SC order enhances the competing SDW order even outside of the normal vortex cores, thus delocalizing magnetic correlations and creating microscopic coexistence of the SDW and SC orders.  The repulsive coupling between SDW and SC orders can be tuned (by chemical substitution or pressure) to tip the balance between the two competing ground states, leading to QPTs among the pure SDW phase, the SDW and SC coexisting phases, and the pure SC phase. 

\section{Conclusion}
Current-voltage measurements were performed at temperatures close to $T_{c0}$ on superconductive \BaCo~single crystals with doping levels covering underdoped, optimally doped, and overdoped regimes. Significantly enhanced flux-flow resistivity was observed at $x=0.06$, possibly related to the existence of the boundary between the purely superconducting phase and the phase where
superconductivity co-exists with the SDW order. The universal scaling behavior of $\rho_{\rm ff}(H,T,x)$ observed in all doping levels implies that the upturn is governed by the dissipation in the bulk excitations. The changes in the scaling parameters $m$, $n$ and $A$ over a wide doping range is in agreement with the fact that changes in the magnetic order around the vortex cores are related to changes in ground state of the system for Co concentrations $x\sim 0.06$. Thus, based on the consistency between the doping level dependence of the upturn in flux-flow resistivity and the phase diagram, we conclude that the abnormal enhancement in $\rho_{\rm ff}$ at low fields is most likely connected to the SDW order around the vortex cores. Our results imply that the viscosity coefficient has strong magnetic field and temperature dependence governed by the systems' proximity to a magnetic instability.

\paragraph{Acknowledgments}
We gratefully acknowledge invaluable help from Alan Baldwin in electronic and software techniques and  Shuai Zhang for helpful discussions and an initial critical reading of the manuscript. This work was supported by the National Science Foundation grants DMR-1505826 and DMR-1506547 at KSU. T. H.  acknowledges the support of NSFC, grant No. 11574338, and XDB04040300. H. X. acknowledges the support of NSFC, grant No. U1530402.

\bibliographystyle{apsrev4-1}
\bibliography{Ba122_Co_IV_2015}


\end{document}